 \documentclass[final,pdftex]{tMOP2e}
  \usepackage{graphicx}
  \graphicspath{{figures/pdf/}}
  \DeclareGraphicsExtensions{.pdf}
\def\vec#1{{\bf#1}}
\def\op#1{\hat{#1}}

\def\norm#1{\| #1 \|}

\def\Tr{\operatorname{Tr}}

\def\diag{\operatorname{diag}}

\def\SU{\mathfrak{SU}}

\def\RR{\mathbb{R}}
\def\NN{\mathbb{N}}
\def\CC{\mathbb{C}}

\def\A{\mathcal{A}}

\def\F{\mathcal{F}}

\def\sx{\op{\sigma}_x}
\def\sy{\op{\sigma}_y}
\def\sz{\op{\sigma}_z}
\def\Had{\mbox{\rm Had}}
\def\CNOT{\mbox{\rm CNOT}}
\def\T{\mbox{\rm T}}
\def\X{\mbox{\rm X}}
\def\Y{\mbox{\rm Y}}
\def\Z{\mbox{\rm Z}}
\def\I{\mbox{\rm I}}
\def\sxyz{\mathbb{\op{\sigma}}}

\begin{document}
\title{Implementation of Quantum Gates via Optimal Control}
\author{Sonia Schirmer$^{a,b,\ast}$%
       \thanks{$^\ast$Corresponding author. Email: sgs29@cam.ac.uk} \\
        $^a$Dept of Applied Maths and Theoretical Physics \\
             University of Cambridge, Wilberforce Road, Cambridge, CB3 0WA, UK \\
        $^b$Dept of Maths and Statistics, University of Kuopio\\
             PO Box 1627, 70211 Kuopio, Finland }
\date{\today}
\maketitle

\begin{abstract}
Starting with the basic control system model often employed in NMR pulse
design, we derive more realistic control system models taking into account 
effects such as off-resonant excitation for systems with fixed inter-qubit
coupling controlled by globally applied electromagnetic fields, as well as
for systems controlled by a combination of a global fields and local control
electrodes.  For both models optimal control is used to find controls that
implement a set of two- and three-qubit gates with fidelity $\ge 99.99\%$.
\end{abstract}


\section{Introduction}

Controlling the dynamics of quantum systems is a problem of great recent
interest with many applications from quantum chemistry to quantum
computing.  Since the dynamics of a quantum system is essentially
determined by its Hamiltonian, a central theme in quantum control is
Hamiltonian engineering.  This is typically accomplished by the
application of external fields, whose interaction with the system
modifies its effective Hamiltonian, thereby changing its dynamical
evolution.  The type of control fields that are available tends to vary
significantly depending on the type of system and the actuators
available.  In traditional applications such as nuclear spin engineering
and nuclear magnetic resonance (NMR), the fields are often
radiofrequency pulses, while in applications involving transitions
between electronic states in atoms or vibrational modes of molecules,
the control fields are usually laser pulses.  In more recent
applications such as various solid-state architectures based on
semi-conductor quantum dots or Josephson junctions, the controls are
often electric fields produced by external electrodes.  Despite the
disparate physical realizations, all of these applications share
important similarities, in that the external fields modify the
Hamiltonian in a certain way, and the objective generally is to find a
particular control that modifies the dynamics in a certain way, and
achieves a certain objective, e.g., of steering the system to a
particular state or realizing a desired quantum process.

There are many differences in the specifics, however, which can be very
important for the design of effective and experimentally feasible
controls.  In NMR applications, for example, we can often approximate
the system well by assuming that the control fields induce rapid local
rotations of individual spins, which interact through a fixed, weak
coupling between them that is not affected by the control fields.
Furthermore, we can often assume that we can selectively address
individual spins by frequency-selective control pulses, and that we can
perform local rotations about orthogonal axes.  The same approximations,
however, are not necessarily valid for other systems such as solid-state
architectures.  In particular for systems controlled by local voltage
gates, it is often difficult to perform rotations about orthogonal axes,
and a voltage applied to a local electrode may have non-negligible
effects on the dynamics of nearby quantum dots, and in the presence of
multiple control voltages there can be signficant cross-talk effects,
all of which can signficantly complicate effective control design.  In
this paper we attempt to incorporate some of these effects to derive
more realistic control system models for two types of systems of
practical interest, and apply optimal control theory to find control
solutions to implement a set of elementary gates for the resulting more
complex control system models.

\section{Basic NMR control system model and geometric control}

The starting point for many optimal control problems in quantum
computing such as optimal implementation of quantum gates has often been
a control system model inspired by liquid-state nuclear magnetic
resonance (NMR).  Using the common shorthand $\op{\sigma}_k^{(n)}$ to
denote an $N$-factor tensor product in which the $n$th factor is the
Pauli matrix $\op{\sigma}_k$ for $k\in \{x,y,z\}$ and all other factors
are the identity $\I$, the Hamiltonian for a system consisting of $N$
spin-$\frac{1}{2}$ nuclei (qubits) in this model is usually given by
\begin{equation}
\label{eq:sys0}
  \op{H}[\vec{u}(t)] 
         = \hbar\sum_{n<n'} J_{nn'} \sz^{(n)} \sz^{(n')}
          +\hbar\sum_{n=1}^N \left(u_{2n-1}(t)\sx^{(n)}
           +u_{2n}(t)\sy^{(n)}\right),
\end{equation}
where the first term is a fixed inter-qubit coupling term, in this case
of Ising type, and we have $2N$ independent control fields $u_{2n-1}(t)$
and $u_{2n}(t)$ capable of performing local $x$ and $y$ rotations on the 
$n$th spin, respectively.  

If the interqubit couplings $J_{nn'}$ are controllable, or at least can
be switched off, then it is very easy to see that we can implement any
local unitary operation $\op{U}\in \SU(2)$ on the $n$th qubit, e.g., by 
performing a sequence of three local rotations about the orthogonal axes 
$\sx^{(n)}$ and $\sy^{(n)}$, e.g., using the standard Euler decomposition 
$\op{U}=\op{U}_x(\alpha)\op{U}_y(\beta)\op{U}_x(\gamma)$ with
$\op{U}_x(\alpha)=\exp(-i\alpha\sx)$, etc, for suitable values of the 
rotation angles $\alpha,\beta,\gamma$, and two-qubit gates can be 
implemented using the Cartan decomposition~\cite{Khaneja01,Zhang,Earp},
\begin{equation}
\label{eq:cartan}
  \op{U} = \op{U}_1 \, 
           [\op{U}_y \, \op{Z}(\alpha_3) \op{U}_y^{-1}]\, 
           [\op{U}_x^{-1}\, \op{Z}(\alpha_2) \, \op{U}_x]\, 
           \op{Z}(\alpha_1) \, \op{U}_2,
\end{equation}
where $\op{Z}(\alpha)=\exp(-i\alpha\sz^{(1)}\sz^{(2)})$ corresponds to
free evolution under the Ising-coupling Hamiltonian for a suitable time,
$\op{U}_1$ and $\op{U}_2$ correspond to local operations on both qubits,
which can be implemented using the Euler decomposition, and
$\op{U}_x=\op{U}_x^{(1)}(\frac{\pi}{4})\otimes\op{U}_x^{(2)}(\frac{\pi}{4})$
correspond to local $x$-rotations by $\frac{\pi}{4}$ on both qubits, and
similarly for $\op{U}_y$.

If the inter-qubit coupling terms $J_{nn'}$ are fixed and not controllable
then this gives rise to a drift term, which complicates the situation.
If the coupling is weak and we can apply ``hard pulses'', e.g., if we
can ensure
\[
  \sqrt{u_{2n-1}^2+u_{2n}^2} \gg \max_{n'} |J_{nn'}|, 
\]
then the drift term is negligible (as far as the implementation of local 
gates is concerned), and the geometric control pulse sequences derived 
from the Euler and Cartan decomposition are often a good approximation.  
However, if the coupling is stronger or hard pulses are not available, 
neglecting the drift term will result in inaccurate gates.  Moreover,
in realistic applications, hard pulses may lead to other problems such
as off-resonant excitation.

Although we can in principle try to find geometric decompositions that 
take the drift term into account~\cite{Ramakrishna}, geometric control 
pulse sequences derived from Lie group decompositions are usually not 
optimal even when the drift is negligible, and optimal control has been 
shown to be able to produce more effective or efficient control pulses 
with regard to various perfomance indices such as gate operation times%
or average fidelity even for cases where geometric decompositions are 
valid, at least theoretically [See e.g.~\cite{Herbruggen}].

\section{Realistic control system models for optimal control}

The success of optimal control theory in finding more effective control 
pulse sequences for various model problems has fueled a surge of interest 
in this area.  However, the control system models on which many of the 
optimal control calculations so far have been based are generally too 
simple and often unrealisitc for the systems they are supposed to model.
The model described in the previous section, for example, is based on 
many approximations that may be questionable even in NMR settings, and 
certainly for other applications.  For instance, in the ideal case the 
control fields $u_{2n-1}(t)$ and $u_{2n}(t)$ induce only an $x$-rotation 
and $y$-rotation, respectively, on the $n$th qubit, but the desire to 
speed up gate implementation times may produce strong control fields 
likely to induce off-resonant excitations not accounted for the in model.  

Off-resonant excitation means that the fields are not only felt by the
target qubit but also by other qubits.  In the simplest case we can try
to model off-resonant excitation by replacing the individual interaction
Hamiltonians $\sx^{(n)}$ and $\sy^{(n)}$ for the independent control
fields $u_{2n-1}(t)$ and $u_{2n}(t)$ by linear combinations of the
coupling terms
\begin{subequations}
\label{eq:lin_xtalk}
\begin{align}
  \sx^{(n)} &\mapsto \op{H}_{2n-1} 
     = \sum_{n'} \alpha_{2n-1,2n'-1}\sx^{(n')}+\alpha_{2n-1, 2n'}\sy^{(n')},\\
  \sy^{(n)} &\mapsto \op{H}_{2n}
     = \sum_{n'} \alpha_{2n,2n'-1}\sx^{(n')}+\alpha_{2n, 2n'}\sy^{(n')}.
\end{align}
\end{subequations}
The new interaction Hamiltonians $\op{H}_m$ for $m=1,\ldots,2N$ no
longer correspond to local rotations about mutually orthogonal axes,
which makes finding geometric decompositions for this case very
challenging in general.  However, even arbitrary (linear) cross-talk
need not be an obstacle for optimal control as shown
in~\cite{Schirmer07} for a model system of this type.  Provided the
cross-talk Hamiltonians are known, based on a combination of theoretical
models and experimental characterization, it still appears to easy to
find effective control pulses using optimal control algorithms.  The
main drawback is that this simple model does not accurately model real
cross-talk.

\subsection{Control via globally applied fields with off-resonant excitation}

The intrisic system Hamiltonian for a system of $N$ spins with resonance 
frequencies $\omega_n$ with static Ising coupling $J_{nn'}$ in a fixed
laboratory frame is
\begin{equation}
 \label{eq:sys1a}
  \op{H}_0 = \hbar \sum_{n<n'} J_{nn'} \sz^{(n)}\sz^{(n')}
            -\hbar \sum_n \frac{\omega_n}{2}\sz^{(n)}. 
\end{equation}
If we apply a global electromagntic field $\vec{B}(t)$ with components 
$B_x(t)=B(t)\cos(\omega t+\phi)$ and $B_y(t)=B(t)\sin(\omega t+\phi)$,
respectively, the resulting control Hamiltonian is
\begin{equation}
 \label{eq:sys1b}
  \op{H}_c = -\frac{\hbar}{2} \sum_{n} \gamma_n B(t)
       [\cos(\omega t+\phi) \sx^{(n)}-\sin(\omega t+\phi) \sy^{(n)}]
\end{equation}
Transforming to a multiply rotating frame defined by
$\op{U}_S(t) = \exp(-i \frac{t}{2}\sum_n \omega_n \sz^{(n)})$, the
interaction picture rotating wave approximation Hamiltonian is~\cite{Vandersypen}
\begin{equation}
  \label{eq:sys1c}
  \op{H} = \hbar \sum_{n,n'} J_{nn'} \sz^{(n)}\cdot\sz^{(n')}
          -\frac{\hbar}{2} \sum_n \gamma_n B(t) 
           [ \cos(\Delta\omega_n t+\phi) \sx^{(n)}
            -\sin(\Delta\omega_n t+\phi) \sy^{(n)} ]
\end{equation}
where $\Delta\omega_n=\omega-\omega_n$ and $\Omega_n(t)=\gamma_n B(t)$.
If the field is exactly resonant with the $n$th spin, $\omega=\omega_n$, 
then the term corresponding to the $n$th spin simplifies to
\begin{equation*}
 -\frac{\hbar}{2} \gamma_n B(t) [\cos(\phi)\sx^{(n)}-\sin(\phi)\sy^{(n)}].
\end{equation*}
If we can further argue that all other spins have frequencies $\omega_n$ 
far detuned from $\omega$, $|B(t)|$ is slowly varying and not too large,
$|\gamma_{n'} B(t)|\ll |\Delta\omega_{n'}|$ for all $n'\neq n$, then 
for sufficiently large $T$ the rapidly oscillating terms will ``average 
to zero'', i.e., we can assume
\begin{align*}
  \frac{1}{T}\int_{0}^T B(t)\gamma_{n'}\cos(\Delta\omega_{n'}t+\phi) \, dt \approx 0 \\
  \frac{1}{T}\int_{0}^T B(t)\gamma_{n'}\sin(\Delta\omega_{n'}t+\phi) \, dt \approx 0.
\end{align*}
In this case the effective control Hamiltonian associated with the field
$\vec{B}(t)$ simplifies to
\[
 \op{H}_c=-\frac{\hbar}{2} \Omega_n(t) [\cos(\phi) \sx^{(n)}-\sin(\phi) \sy^{(n)}].
\]
Thus, if we are able to simultaneously apply $N$ independent fields
$B_x^{(n)}(t)\cos(\omega_n t)$ with frequencies $\omega_n$ and $N$
independent fields $B_y^{(n)}(t)\sin(\omega_n t)$ with frequencies
$\omega_n$, for instance, then assuming all the previous assumptions and
approximations still hold, we obtain
\begin{equation}
  \op{H} = \hbar \sum_{n,n'} J_{nn'} \sz^{(n)}\cdot\sz^{(n')}
          -\frac{\hbar}{2} \sum_n [B_x^{(n)}(t)\gamma_n \sx^{(n)}
                                  -B_y^{(n)}(t)\gamma_n \sy^{(n)}],
\end{equation}
exactly the Hamiltonian~(\ref{eq:sys0}) with
$u_{2n-1}(t)=-\frac{1}{2}\gamma_n B_x^{(n)}(t)$ and 
$u_{2n}(t)=\frac{1}{2}\gamma_n B_y^{(n)}(t)$.  

In practice, it is questionable, however, whether these approximations
are still valid in general, especially if we simultaneously apply $2N$
control fields $B_x^{(n)}(t)$ and $B_y^{(n)}(t)$, all of which may be
strongly modulated in time as a result of an optimization procedure, and
thus not slowly varying.  Furthermore, the interaction picture RWA
Hamiltonian~(\ref{eq:sys1c}) is not a very good starting point for a
general optimization either because each of the independent control
Hamiltonians
\begin{equation}
 \op{H}_m(t) =
  -\frac{\hbar}{2} \sum_n \gamma_n 
           [ \cos(\Delta\omega_{n,m} t+\phi_m) \sx^{(n)}
            -\sin(\Delta\omega_{n,m} t+\phi_m) \sy^{(n)} ]
\end{equation}
with $\omega_{n,m}=\omega_m-\omega_n$ for a field $\vec{B}_m(t)$ with 
$B_x(t)=B_m(t)\cos(\omega_m t +\phi_m)$ and 
$B_y(t)=B_m(t)\sin(\omega_m t +\phi_m)$, in the total Hamiltonian 
\begin{equation}
  \op{H} = \op{H}_0 + \sum_{m} B_m(t) \op{H}_m(t)
\end{equation}
is time-dependent, containing rapidly oscillating sine and cosine terms.
Therefore, we might as well let the component functions $B_x(t)$ and
$B_y(t)$ of the field be completely arbitrary and work with the general 
Hamiltonian
\begin{equation}
 \label{eq:sys1}
  \op{H} = \op{H}_0 - B_x(t) \frac{\hbar}{2} \sum_{n} \gamma_n \sx^{(n)}
                    + B_y(t) \frac{\hbar}{2} \sum_{n} \gamma_n \sy^{(n)}
\end{equation}
with $\op{H}_0$ as in Eq.~(\ref{eq:sys1a}), instead.

\subsection{Control via fixed global fields and local control electrodes}

In some applications such as NMR the only type of control we have is
globally applied control fields.  However, for many systems, especially
solid-state systems such as the nuclear spins of donor atoms embedded in
a substrate~\cite{Kane}, or electron spins in quantum dots~\cite{Loss}, 
for example, some local control in the form of control electrodes is 
available, and in a typical setting, control is achieved by combining 
globally applied electromagnetic fields with local fields provides by 
control electrodes~\cite{Hill}.  In this case we often have a global 
electromagnetic field $\vec{B}(t)$ with components 
$B_x(t)=B(t)\cos(\omega t)$, $B_y(t)=B(t)\sin(\omega t)$, 
in addition to a static magnetic field in the $z$ direction, and a
Hamiltonian of the form
\begin{equation}
 \label{eq:sys2a}
  \begin{split}
  \op{H} =& \hbar \sum_{n<n'} J_{nn'} \sxyz^{(n)}\cdot\sxyz^{(n')}
            -\frac{\hbar}{2} \sum_n  \omega_n(\vec{V}) \sz^{(n)}\\
          & -\frac{\hbar}{2} \sum_{n} \gamma_n B(t)
             [\cos(\omega t) \sx^{(n)}-\sin(\omega t) \sy^{(n)}],
  \end{split}
\end{equation}
as before, but in addition we now have some degree of control over the 
resonance frequency $\omega_n$ of the $n$th qubit via local control 
electrodes.  Note that unlike in the previous model we have assumed 
isotropic Heisenberg coupling, i.e., $\sxyz^{(n)}\cdot\sxyz^{(n')}
=\sx^{(n)}\sx^{(n')}+\sy^{(n)}\sy^{(n')}+\sz^{(n)}\sz^{(n')}$, instead
of $\sz^{(n)}\sz^{(n')}$ Ising coupling as this seems to be more
appropriate for solid-state systems, although the assumption of 
isotropy may not be appropriate for some systems. 

One control approach is to simultaneously vary the voltages applied to
the local control electrodes as well as the globally applied field.
However, it practice it is much simpler to apply a global field
$\vec{B}(t)$ with a fixed amplitude $B$ and fixed frequency $\omega$ and
only vary the voltages applied to the local control electrodes.  In this
case we can introduce a rotating frame, this time rotating at the field
frequency $\omega$,
$\op{U}_S(t)=\exp(-i\frac{t}{2}\sum_n\omega\sz^{(n)})$, to obtain the
slightly simpler interaction picture Hamiltonian
\begin{equation}
\label{eq:sys2}
  \op{H} = \hbar\sum_{n<n'} J_{nn'} \sxyz^{(n)} \cdot \sxyz^{(n')}
           -\frac{\hbar}{2} \sum_n \omega_n(\vec{V}) \sz^{(n)}
           -\hbar\Omega \sum_{n} \bar\gamma_n\sx^{(n)},
\end{equation} 
where we set $\Omega=\frac{1}{2}B\gamma_0$ and take $\bar\gamma_n$ to be
the coupling constants $\gamma_n$ in units of some coupling strength
$\gamma_0$, so that for a homogeneous system consisting of a chain of
spins that couple identically to the field we have $\bar\gamma_n=1$ for
$n>0$.

In this model the constants $J_{nn'}$ and $\gamma_n$ are usually fixed
and determined by the physics of the system.  The frequency $\omega$ and
Rabi frequency $\Omega$ of the field can be chosen initially but are
then fixed.  Thus, all the dynamic control is achieved by dynamically
changing the control voltages and thus $\omega_n(\vec{V})$.  This type
of control therefore leads to a different partitioning of the total
Hamiltonian into a fixed drift and variable control part
\begin{align*}
  \op{H}_0 &=  \hbar \sum_{n<n'} J_{nn'} \sxyz^{(n)}\cdot\sxyz^{(n')}
              -\hbar \Omega \sum_{n} \bar\gamma_n \sx^{(n)}, \\
  \op{H}_c &= -\frac{\hbar}{2} \sum_n \omega_n(\vec{V}) \sz^{(n)}.
\end{align*}
In general, due to cross-talk effects~\cite{Kandasamy} the energy level
shift of the $n$th qubit may be a complicated function of all the
voltages applied, not only to the electrode directly above it, but to
neighbouring control electrodes as well, i.e., $\omega_n$ may be a
function of all the control voltages $\vec{V}=(V_1,\ldots,V_m)$ applied,
i.e., $\omega_n=\omega_n(\vec{V})$. 

\section{Optimal gate implementation}

\subsection{Optimal control formulation}

One way to solve the problem of finding controls $\vec{u}(t)$ that
implement a desired set of quantum gates, especially for control systems
that are not amenable to simple geometric decomposition schemes, such as
the ones discussed in the previous section, is to formulate the task as
an optimization problem.  This typically involves (a) choosing a target
function to be optimized, (b) finding a suitable parametrization for the 
set of admissible (and experimentally feasible) controls, and (c) finding 
a solution to the resulting, possibly constrained, optimization problem
based on application of Pontryagin's maximum principle~\cite{Pontryagin}. 

For the task of implementation of quantum gates or processes, a natural 
choice for the objective function is the gate fidelity~\cite{Nielsen}
\begin{equation}
  \F(\op{U}_T,\op{U}) = \frac{1}{N}\Re\Tr(\op{U}_T^\dagger\op{U}).
\end{equation}
The reason for the factor of $1/N$, where $N$ is the system dimension, 
is to ensure that $\F$ varies between $-1$ (for $\op{U}=-\op{U}_T$) and 
$+1$ (for $\op{U}=\op{U}_T$) independent of the system dimension.  In
general, maximizing the gate fidelity is equivalent to minmizing the 
gate error
\begin{equation}
  \mathcal{E}(\op{U}_T,\op{U}) = \norm{\op{U}-\op{U}_T}^2.
\end{equation}
This is very to see if the target operator $\op{U}_T$ and the 
actual operator $\op{U}$ are both unitary (and thus
$\op{U}^\dagger\op{U}=\op{U}_T^\dagger\op{U}_T=\op{I}$) as
\begin{equation*}
\begin{split}
  \norm{\op{U}-\op{U}_T}^2 
  &= \Tr[(\op{U}-\op{U}_T)^\dagger(\op{U}-\op{U}_T)] \\ 
  &= \Tr[\op{U}^\dagger\op{U} + \op{U}_T^\dagger\op{U}_T
        -\op{U}^\dagger\op{U}_T -\op{U}_T^\dagger \op{U}] \\ 
  &= 2N - 2\Re\Tr(\op{U}_T^\dagger\op{U}).
\end{split}
\end{equation*}
Noting that the implemented operator $\op{U}=\op{U}_{\vec{u}}(t_F)$ is a
function of the control $\vec{u}$, the control objective is to find an 
admissible control $\vec{u}_{\rm opt}$ such that
\begin{equation}
  \F(\op{U}_T,\op{U}_{\vec{u}_{\rm opt}}(t_F))
   = \max_{\vec{u}\in\A} \F(\op{U}_T,\op{U}_{\vec{u}}(t_F)),
\end{equation}
and of course, $\op{U}_{\vec{u}}(t)$ has to satisfy the Schrodinger equation
\begin{equation}
  \frac{d}{dt}\op{U}_{\vec{u}}(t)
  = -\frac{i}{\hbar} \op{H}[\vec{u}(t)] \op{U}_{\vec{u}}(t),
\end{equation}
where $\op{H}[\vec{u}(t)]$ is the control-dependent Hamiltonian of the
system.

\subsection{Iterative control optimization algorithm}

Once the Hamiltonian $\op{H}[\vec{u}(t)]$, the target time $t_F$ and the
target operator $\op{U}_T$ have been chosen, the fidelity depends only on
the choice of the control $\vec{u}(t)$, i.e., $\F=\F_{\vec{u}}$  Now let 
$\vec{u}$ and $\vec{u}+\Delta\vec{u}$ be two controls.  Noting that
\begin{equation}
  \delta\op{U}_{\vec{u}}(t)
  := \op{U}_{\vec{u}+\Delta\vec{u}}(t)-\op{U}_{\vec{u}}(t)
  = \int_{t_0}^{t}\op{U}_{\vec{u}}(t,\tau)\op{D}_{\Delta\vec{u}}(\tau)\, d\tau
\end{equation}
and that for a control-linear Hamiltonian 
$\op{H}[\vec{u}(t)]=\op{H}_0+\sum_m u_m(t) \op{H}_m$ we have
\begin{equation}
   \op{D}_{\Delta\vec{u}}(t) 
 =-\frac{i}{\hbar} \sum_m \Delta u_m(t) \op{H}_m \op{U}_{\vec{u}+\Delta\vec{u}}(t)
\end{equation}
we obtain immediately
\begin{align*}
\Delta\F 
&= \F_{\vec{u}+\Delta\vec{u}} -\F_{\vec{u}} \\
&= \frac{2}{N} \Re \Tr[\op{U}_T^\dagger 
    (\op{U}_{\vec{u}+\Delta\vec{u}}(t) - \op{U}_{\vec{u}}(t)) ]\\
&= \frac{2}{N\hbar} 
    \Im \Tr \left[\op{U}_T^\dagger\int_{t_0}^{t_F}
            \op{U}_{\vec{u}}(t_F,t)
    \sum_m \Delta u_m(t) \op{H}_m \op{U}_{\vec{u}+\Delta\vec{u}}(t)\, dt \right] \\
&= \frac{2}{N\hbar} \sum_m \int_{t_0}^{t_F} \Delta u_m(t) 
    \Im \Tr \left[\op{U}_T^\dagger
            \op{U}_{\vec{u}}(t_F,t)
            \op{H}_m \op{U}_{\vec{u}+\Delta\vec{u}}(t)\right] \, dt 
\end{align*}
which shows that setting
\begin{equation}
 \label{eq:Deltau}
  \Delta u_m(t) = \epsilon_m(t) \Im \Tr \left[\op{U}_T^\dagger
            \op{U}_{\vec{u}}(t_F,t)
            \op{H}_m \op{U}_{\vec{u}+\Delta\vec{u}}(t)\right] 
\end{equation}
with $\epsilon_m(t)>0$ for all $t$ will increase the fidelity.  Thus,
starting with any initial guess $\vec{u}_0(t)$, iteratively solving 
the Schrodinger equation while updating the control in each iteration 
according to the rule $\vec{u}^{(n+1)}=\vec{u}^{(n)}+\Delta\vec{u}^{(n)}$,
with $\Delta\vec{u}^{(n)}$ chosen as in Eq.~(\ref{eq:Deltau}), will 
monotonically increase the fidelity in each iteration, and as the 
fidelity is of course bounded, the iteration must converge to some
limiting fidelity $\F^\infty$, although we cannot guarantee that the
limiting value will be equal to the global maximum of the fidelity.
In fact, many optimal control algorithms are based precisely on this
iterative update scheme.  

\subsection{Parametrization of controls and discretization}

To solve the Schrodinger equation numerically, we still require some
discretization.  The simplest approach is to assume piecewise constant
fields $\vec{u}(t)=\vec{u}_k$ for $t_{k-1}\le t\le t_k$, in which case 
the exact solution of the Schrodinger equation is explicitly
\begin{equation}
  \op{U}_{\vec{u}}(t_F)
 = \op{U}_{\vec{u}_K}(\Delta t_K) \op{U}_{\vec{u}_{K-1}}(\Delta t_{K-1}) 
   \ldots \op{U}_{\vec{u}_1}(\Delta t_1) 
\end{equation}
where $\Delta t_k= t_k-t_{k-1}$, $t_K=t_F$, and
\begin{equation}
  \op{U}_{\vec{u}_k}(\Delta t_k)
  = \exp \left[ -\frac{i}{\hbar} \Delta t_k \op{H}(\vec{u}_k) \right]
\end{equation}
is a simple matrix exponential and the update rule becomes
\begin{equation}
 \label{eq:Deltau2}
  \Delta u_m(t_k) = 
  \epsilon_m(t) \Im \Tr \left[\op{U}_T^\dagger
  \op{U}_{\vec{u}_{K}}(t_K) \cdots \op{U}_{\vec{u}_{k+1}}(t_{k+1})\op{H}_m  
  \op{U}_{\vec{u}_k}(\Delta t_k) \cdots \op{U}_{\vec{u}_1}(\Delta t_1) \right]. 
\end{equation}
With this type of discretization there are two ways of implementing the
update rule.  In the global update approach all the matrix exponentials
for all times $t_k$ are evaluated to solve the Schrodinger equation, and 
the fields are updated for all times $t_k$ simultaneously at the end of 
each iteration step.  This is essentially the well-known GRAPE algorithm
as described in~\cite{Khaneja}. An alternative is to update the fields 
$u_m(t_k)$ at each time step $t_k$.  Our analysis above shows that both 
approaches will monotonically increase the fidelity provided $\epsilon_m(t_k)>0$ 
for all $t_k$, although they will tend to converge to different 
solutions for the fields.  

In our limited numerical experiments for the systems considered here,
both approaches seemed equally effective in finding satisfactory control
fields but the local update approach tended to converge faster in terms
of the computational time required to find a solution.  The main reason
for this appears to be that in the local update approach only a single
$\vec{u}_k$ is incremented at a time, and hence we can optimize the
multiplier $\epsilon_k$ with very little computational overhead to speed 
up the convergence, as any local change of $\epsilon_k$ requires only
re-evaluation of a single matrix exponential (and two matrix
multiplications and a trace operation)
to assess its effect on the fidelity, while in the global update any
change in $\epsilon$ requires the re-evaluation of all the $K$ matrix 
exponentials and $K-1$ matrix multiplications to assess the fidelity of
the new control.  An alternative to this simple step-size control is to
use higher derivatives to predict the optimal step-size but this also
involves potentially significant computational overhead to find the best
step size, in addition to the need to calculate analytical expressions 
for these derivates, thus it is not clear if this will yield a 
computational advantage or not, although it is worth investigating.

A general shortcoming of this simple optimization procedure is that it
does not take constraints into account or include penalities for less
desirable fields.  Simple constraints such as upper bounds on the field
amplitudes can easily be incorporated into the algorithm.  More complex
constraints such as limiting the frequency bandwidth of pulses or
minimizing the overall pulse energies etc require augmentation of the
objective functional by constraint or penalty terms and modifications to
the update rules.  For some types of penalty terms iterative algorithms
very similar to the algorithm above exist [see e.g.~\cite{Schirmer00}], 
but a detailed discussion of appropriate penalty functions, resulting 
iterative update rules, and comparison of the resulting optimal fields
is beyond the scope of this paper.

\section{Simulation details and results}

The specific task we wish to accomplish here is finding controls 
$\vec{u}(t)$ that implement the following complete set of universal 
gates (i.e., target operators)
\begin{equation} 
\label{eq:UT1}
 \op{U}_T \in \{\I\otimes\I,\Had\otimes\I, \T\otimes\I, 
 \I\otimes\Had, \I\otimes\T, \CNOT\}
\end{equation} 
for a two-qubit system, where we employ the common appreviations
$\Had=\exp(i \frac{\pi}{4} Y)$, $T=\exp(i\frac{\pi}{8} Z)$ and
$\CNOT=e^{-i\pi/4}\diag(I,X)$.  The phase factor for the $\CNOT$ is
necessary to ensure that the target operators are in $\SU(4)$ since our
model Hamiltonians have zero trace and hence can only generate operators
in $\SU(4)$.  As a second, more challenging task we consider the
implementation of the Toffoli-like three-qubit gate
\begin{equation}
\label{eq:UT2}
  \op{U}_T = \diag(1,1,1,1,1,1,iX) \in \SU(8).
\end{equation}

For the simulations we assume uniform nearest-neighbour coupling, i.e.,
$J_{nn'}=J\delta_{n,n'-1}$.  Since the coupling constants $J$ in our
model have units of frequency, by choosing time in units of $1/J$, we
can set the coupling constants to unity.  All other frequencies, unless
otherwise states, are also chosen in units of $J$.  We can furthermore
choose the units of energy such that $\hbar$, which has units of energy
$\times$ time, equals $1$, and can thus be omitted.  For the two-qubit 
gates we set the target gate operation time to $t_F=1$, while for the 
three-qubit gate we choose $t_F=5$ as a Toffoli gate normally requires 
five two-qubit gates.  We choose these values as reasonable target times 
although we were able to find solutions for smaller $t_F$ for many gates, 
suggesting that the gate operation times could in fact be improved.  

\subsection{Global control fields with off-resonant excitation}

For our first model system with Hamiltonian~(\ref{eq:sys1}), assuming 
$\gamma_n=\gamma_0$ for all $n$, and setting 
$u_1(t)=-\frac{1}{2}\gamma_0 B_x(t)$ and
$u_2(t)= \frac{1}{2}\gamma_0 B_y(t)$, the Hamiltonian becomes
\begin{equation}
 \label{eq:model1:2q}
  \op{H} = \Z\Z - \frac{1}{2}(\omega_1\Z\I +\omega_2\I\Z) 
                 + u_1(t) (\X\I+\I\X) + u_2(t) (\Y\I + \I\Y)
\end{equation}
for a two-qubit system, and for a three-qubit system we have 
\begin{equation}
 \label{eq:model1:3q}
  \begin{split}
  \op{H}
   = & (\Z\Z\I+\I\Z\Z) 
  - \frac{1}{2}(\omega_1\Z\I\I +\omega_2\I\Z\I +\omega_3\I\I\Z) \\
     & + u_1(t) (\X\I\I+\I\X\I+\I\I\X) + u_1(t) (\Y\I\I+\I\Y\I+\I\I\Y).
\end{split}
\end{equation}

\begin{figure}
\scalebox{0.7}{\includegraphics{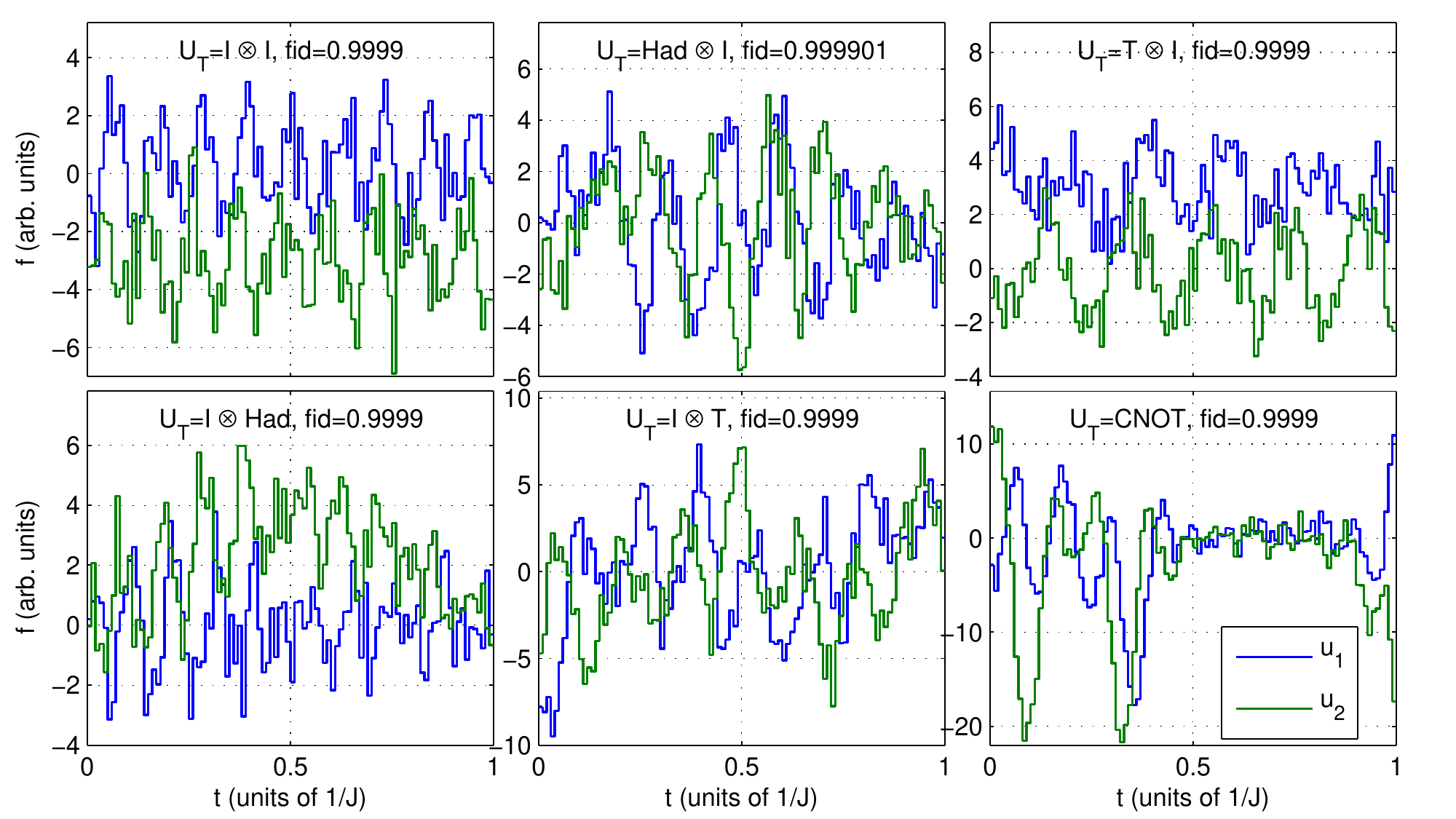}}
\caption{Control field solutions for system~(\ref{eq:model1:2q}) 
with $\omega_1=10$ and $\omega_2=12$ for elementary gates~(\ref{eq:UT1}) 
and $t_f=1$ with $100$ time steps and no field constraints.}
\label{fig1}
\end{figure}

Fig.~\ref{fig1} shows the actual control field solutions for a system
with $\omega_1=10$ and $\omega_2=12$, i.e., a system with Ising coupling
frequency about 10\% of the smaller qubit frequency, i.e., strong enough
to be non-negligible for single qubit operations, and a qubit frequency
difference large enough to allow some selectivity but still small enough
so that off-resonant excitation is a concern for pulses of the magnitude
and duration required for to achieve the desired gate operation times.
For this model system we choose a relatively large number of time steps,
$K=100$, motivated by the assumption that the $x$ and $y$ components of
the globally applied electromagnetic field could be varied more or less
arbitarily, although simulations suggest that we can find piecewise
constant solutions with much fewer steps, if necessary.

The solutions are not unique and in fact different initial fields or
update rules generally produce different solutions.  Again, despite the
increased complexity of the model (off-resonant excitation, no RWA) and
the reduced number of control fields (two vs four in the simpler
model~(\ref{eq:sys0})), it appeared easy to find solutions that achieved
the target fidelity of $\ge0.9999$, or gate errors $\le 0.01$\%.
However, as the figure shows, the magnitudes of the fields for at least
some of the gates are relatively large.  Further analysis also shows
that the spectral range of the optimal controls produced by this
algorithm tends to be rather broad, which may be a concern in practice
when there are bandwidth limitations as filtering of high or low
frequency components will reduce the gate fidelity.  Some numerical
simulations for related problems suggest that solutions with much lower
spectral bandwidth often exist but it appears that a different
parametrization of the fields and optimization approach are required to
find such solutions.  Finding controls to implement the three-qubit gate
proved more of a challenge for this system, in part because convergence
was extremely slow.  Fig.~\ref{fig}(left) shows a solution with fidelity 
$\ge 99.99$\%.  Notice that the fields have high amplitudes and are very 
noisy.  Fig.~\ref{fig4}(left) also shows that the fields have very large 
spectral bandwidth as expected, which would make faithful implementation 
in the laboratory challenging.

\subsection{Control via fixed global fields and local control electrodes}

For our second model system with Hamiltonian~(\ref{eq:sys2}), assuming 
again $\gamma_n=\gamma_0$ for all $n$, and setting 
$u_n(t)=-\frac{1}{2}\omega_n(\vec{V}(t))$ for $n=1,\ldots,N$, the 
Hamiltonian becomes
\begin{equation}
 \label{eq:model2:2q}
  \op{H} = \X\X+\Y\Y+\Z\Z-\Omega (\X\I+\I\X) + u_1(t)\Z\I + u_2(t)\I\Z
\end{equation}
for a two-qubit system, and for a three-qubit system we have
\begin{equation}
 \label{eq:model2:3q}
 \begin{split}
  \op{H} =& (\X\X\I+\Y\Y\I+\Z\Z\I+\I\X\X+\I\Y\Y+\I\Z\Z) -\Omega (\X\I\I+\I\X\I+\I\I\X) \\
          &  + u_1(t)\Z\I\I + u_2(t)\I\Z\I + u_3\I\I\Z.
 \end{split}
\end{equation}
Notice that we take $u_n(t)$ to be independent controls here to
circumvent the complex and highly architecture specific problem of
mapping the actual control voltages $\vec{V}$ onto the frequency shifts
$\omega_n(\vec{V})$, although it must be stressed that optimal control
solutions for $\vec{u}(t)$ will only be useful for a specific system if 
the $\vec{u}(t)$ can be actually realized through proper choice of the 
voltages, which is likely to impose additional, non-trivial constraints.

\begin{figure}
\scalebox{0.7}{\includegraphics{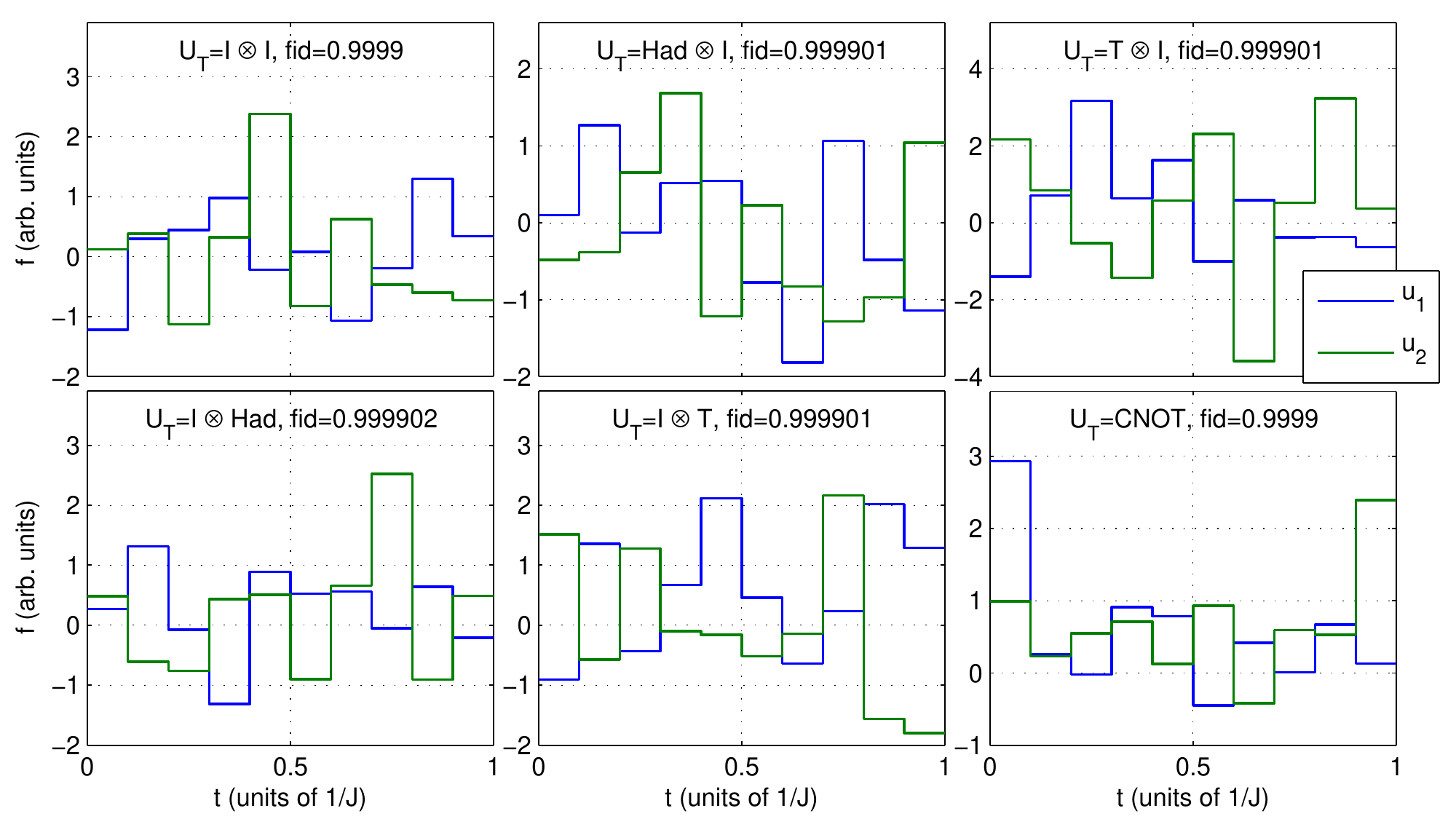}}
\caption{Control field solutions for system~(\ref{eq:model2:2q}) with
$\Omega=10$ for elementary gates~(\ref{eq:UT1}) and $t_f=1$ with $10$
 time steps and no field constraints.}
\label{fig2}
\end{figure}

\begin{figure}
\scalebox{0.45}{\includegraphics{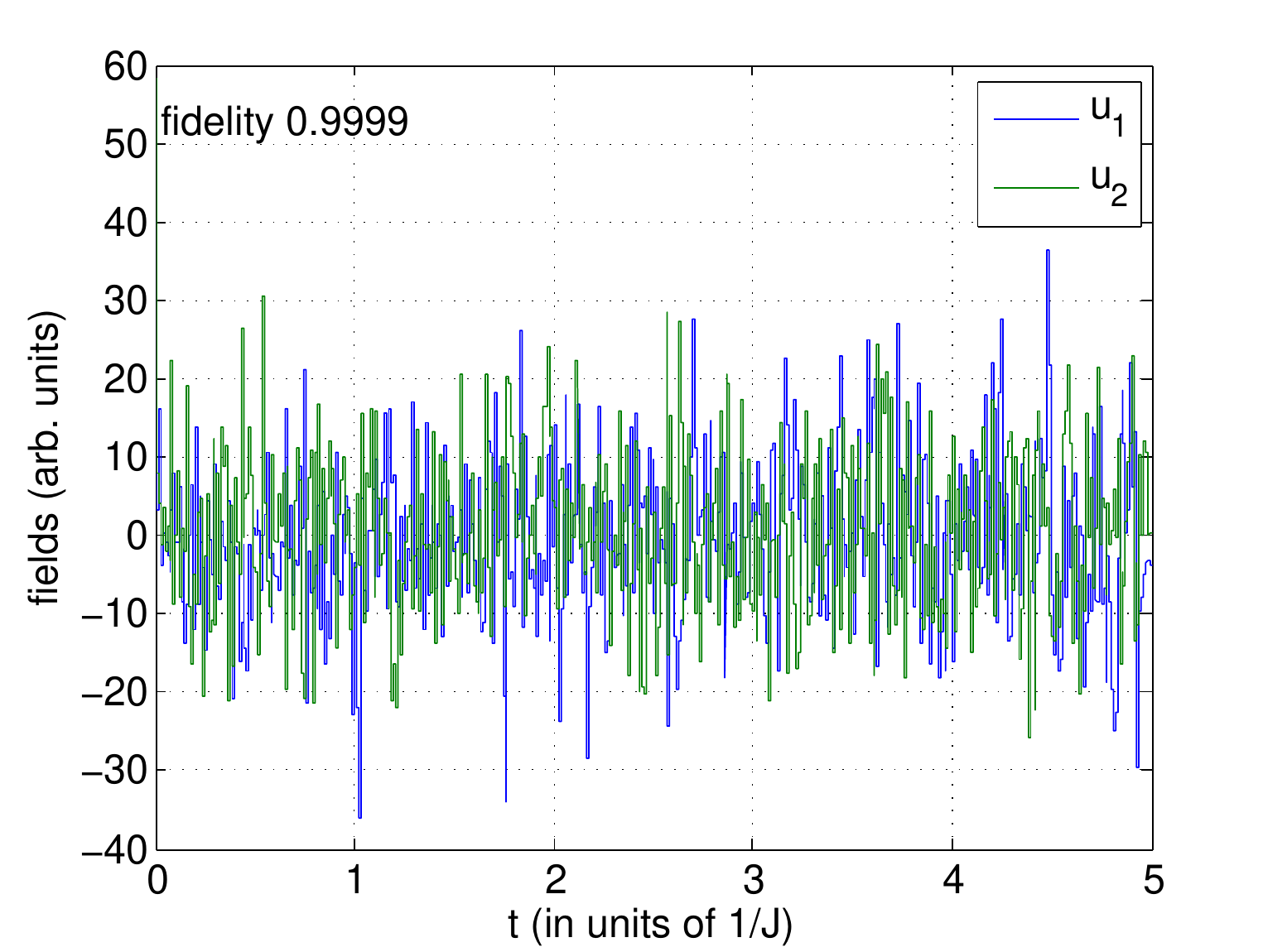}}
\scalebox{0.45}{\includegraphics{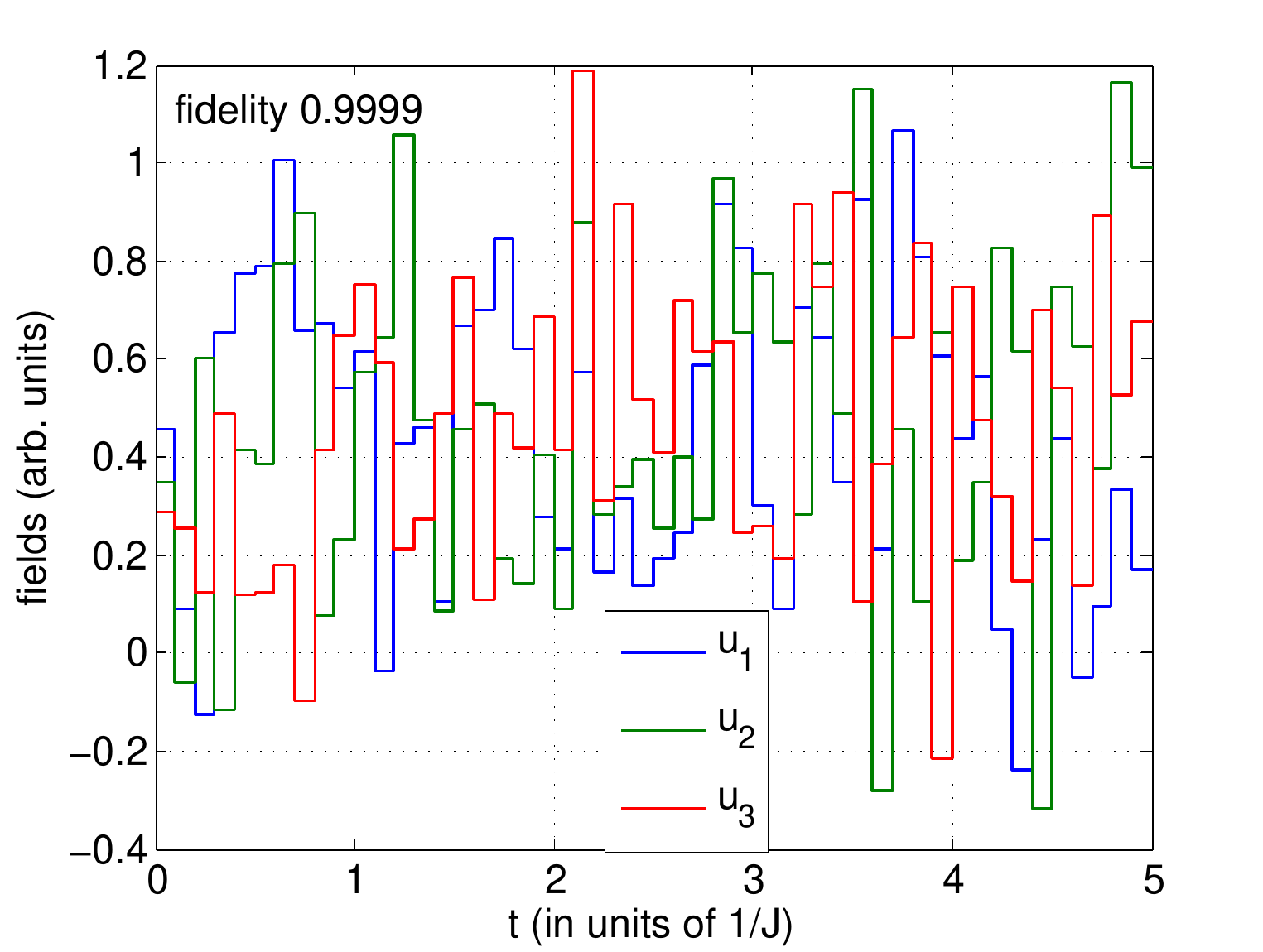}}
\caption{Control field solutions for three-qubit Toffoli-like gate
for model system~(\ref{eq:model1:3q}) with
$\omega_1=10$, $\omega_2=12$ and $\omega_3=8$ (left),
and model system~(\ref{eq:model2:3q}) with $\Omega=10$ (right).}
\label{fig3}
\end{figure}

\begin{figure}
\scalebox{0.45}{\includegraphics{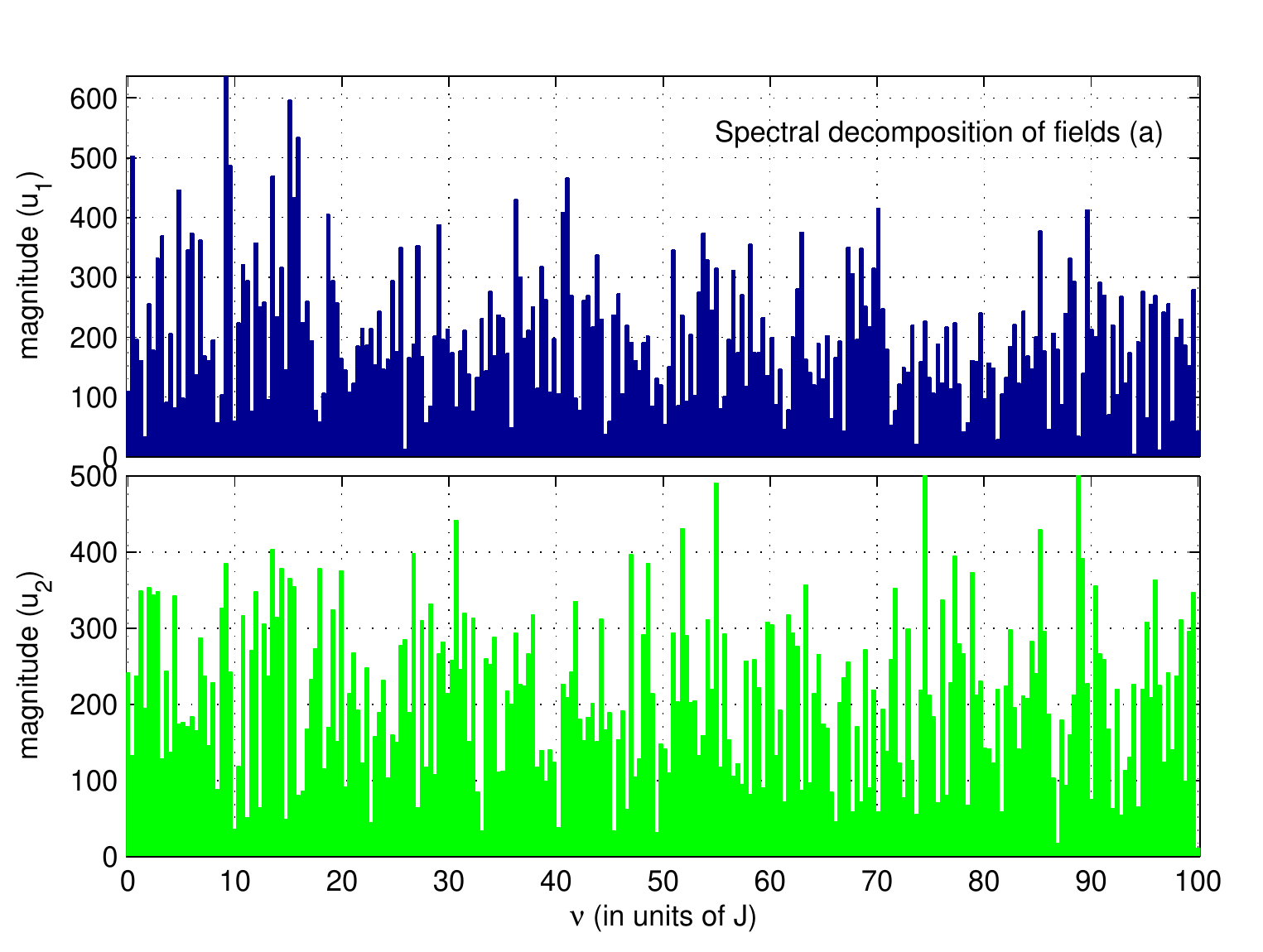}}
\scalebox{0.45}{\includegraphics{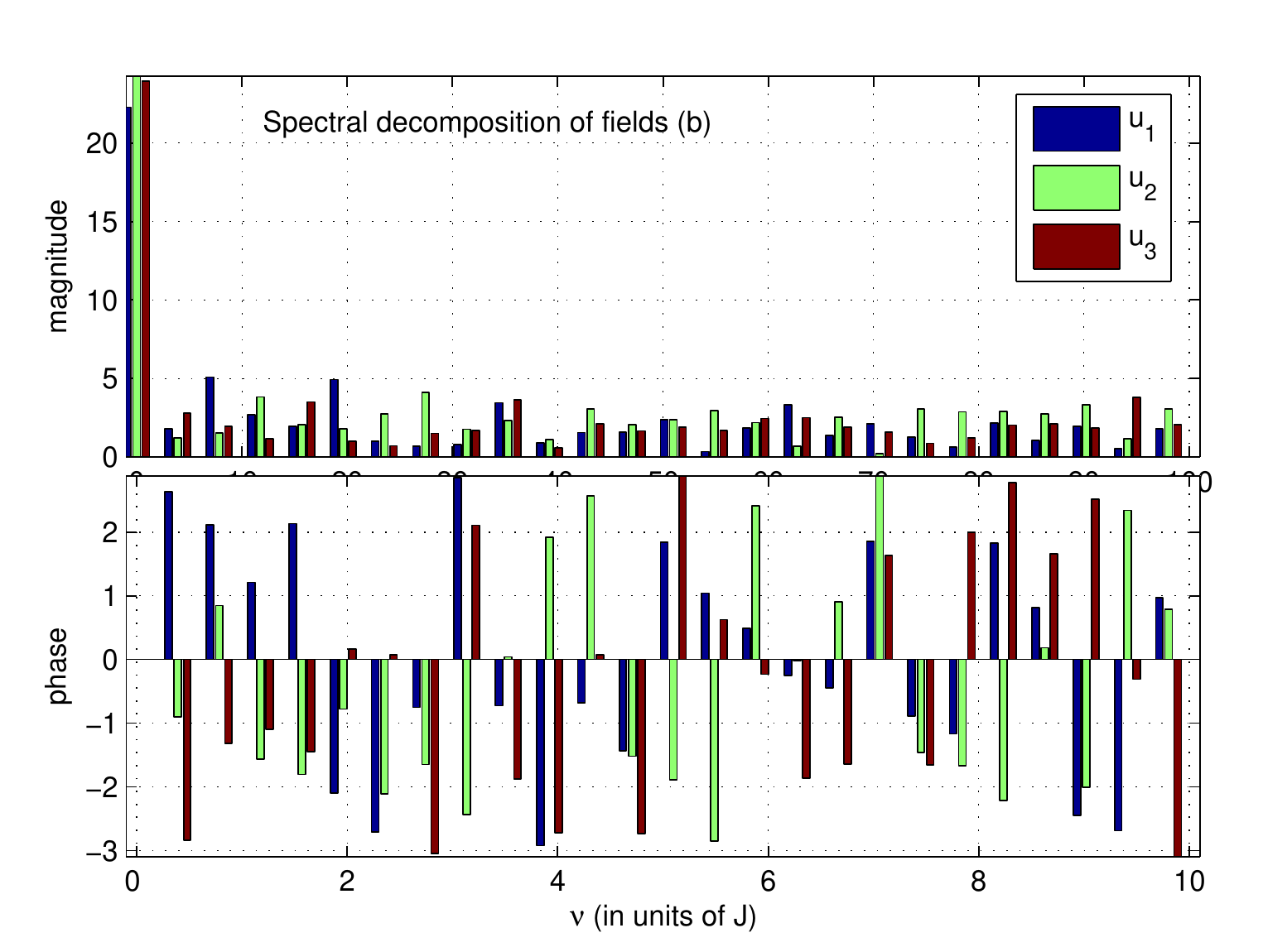}}
\caption{Spectrum of control fields for three-qubit Toffoli-like gate
for model system~(\ref{eq:model1:3q}) with
$\omega_1=10$, $\omega_2=12$ and $\omega_3=8$ (left),
and model system~(\ref{eq:model2:3q}) with $\Omega=10$ (right).}
\label{fig4}
\end{figure}

Again, we chose target gate operation times of $t_F=1$ and $t_F=5$ for
the two- and three-qubit gates, respectively, but in this case we tried
to minimize the number of steps $K$, assuming that frequent switching of
the control voltages would be experimentally more challenging and
increase errors.  As Fig.~\ref{fig2} shows for $\Omega=10$ (in units of
$J$) we were able to find satisfactory solutions for all six two-qubit
gates with only $K=10$ switches.  We note that $\Omega\approx 10$
appeared to be optimal in the sense that we had to increase $K$
significantly for larger or smaller values of $\Omega$.  For $\Omega=1$,
for example, we had to increase the number of steps to $K=40$ and double
the target time to be able to implement the first five two-qubit gates
(corresponding to local operations) with fidelity $\ge 0.9999$, and for
the final CNOT gate the fidelity we achieved was still only $0.9996$,
i.e., below threshold.  For $\Omega=5$ we were able to find solutions
above threshold fidelity for all target gates for $K=20$; for $K=10$ the
fidelity of the CNOT gate remained below threshold ($99.33$\%).
Similarly, for $\Omega=20$ we were able to find solutions above the
threshold fidelity for all gates for $K=20$ but not for $K=10$.  For
even larger $\Omega$, such as $30$ or $40$, $K$ had to be increased to
$30$, and for $\Omega$ in the 50-60 range $K=50$ steps were required.
These results can be partially explained considering that if $\Omega$ is
on the order of the fixed coupling strengths then the weight of the
drift term makes it harder to implement local operations hence the
increase in the time required.  For large $\Omega$ local operations
become easier to implement but entangling gates such as the CNOT, which
rely on generation of entanglment via the fixed Ising coupling term,
become more challenging, requiring more complex pulse shapes, hence
necessitating a larger number of steps $K$.  It is interesting to note
that for this system we had no trouble finding solutions to implement
the three-qubit gate with fidelity $\ge 99.99$\%.  A possible solution 
is shown in Fig.~\ref{fig3}(right).  Although the pulse profiles appear 
more complicated, their magnitudes are actually lower than for some 
of the two-qubit gates, and although $K=50$, the number of switches 
per time unit is still only $10$.  Unlike for the previous model
Fig.~\ref{fig4}(right) shows that the spectral profile and bandwidth 
of the solutions are much more desirable.

\section{Conclusion}

Starting with the most basic control system model commonly used in
applications such as (liquid-state) NMR, we have derived more realistic
control system models for two types of control of special interest for
many applications: (a) control of qubits with fixed inter-qubit coupling
using globally applied electromagnetic fields in the regime where
off-resonant excitation by the control fields is non-negligible, and (b)
control of qubits with fixed coupling using a combination of a fixed
globally applied control field and local control electrodes to shift the
resonance frequencies of the qubits.  Despite the substantially
increased complexity of the resulting models, reduced number of
independent controls, and non-trivial constraints on the allowed
frequency of voltage changes in the latter model, optimal control
techniques even in the rather basic form implemented, overall still
proved very effective in finding solutions that achieved the target
fidelity of $99.99$\% for almost all cases, although there appears to be
scope for further improving the controls with regard to amplitude and
bandwidth considerations, and for improving the overall convergence of
the algorithm, which even with some of the modifications discussed,
appears to be rather slow especially for the three-qubit gates.  A
somewhat unexpected conclusion of the calculations is that our second
model system, which is controlled mainly by local voltage gates that can
only alter the detuning of the qubits from the fixed (unmodulated)
global field appears to be easier to control, even when constraints on
the number of voltage changes are imposed, than the first model system,
which is controlled by global fields with arbitrary pulse envelopes.

\section{Acknowledgements}

SGS acknowledges UK research council funding from an \mbox{EPSRC} 
Advanced Research Fellowship and additional support from the 
\mbox{EPSRC QIP IRC} and Hitachi.  She is currently also a Marie 
Curie Fellow under the European Union Knowledge Transfer Programme 
MTDK-CT-2004-509223.

\end{document}